\documentclass[prl,twocolumn,amsmath,amssymb]{revtex4}

\usepackage{graphicx,graphics,wrapfig,rotating}         
\usepackage{bm,fancybox}
\usepackage{amsmath,amssymb}                          
\usepackage{times,euscript,oldgerm}              %
\usepackage[english]{babel}                      %
\usepackage{psfrag}
\usepackage{color}
\usepackage{epsfig}
\usepackage{dcolumn}

\newcommand{\Yb}{\ensuremath{^{171}\mathrm{Yb}^+}}

\newcommand{\up}{\ensuremath{\left|\uparrow\right\rangle}}
\newcommand{\dn}{\ensuremath{\left|\downarrow\right\rangle}}
\newcommand{\upy}{\ensuremath{\left|\uparrow_y\right\rangle}}

\newcommand{\upz}{\ensuremath{\left|\uparrow_z\right\rangle}}
\newcommand{\dnz}{\ensuremath{\left|\downarrow_z\right\rangle}}

\newcommand{\ket}[1]{|\!#1 \rangle }

\begin{document}

\title{Emergence and Frustration of Magnetic Order with \\ Variable-Range Interactions in a Trapped Ion Quantum Simulator}
\author{R. Islam$^{1 \dag}$, C. Senko$^1$, W. C. Campbell$^{1 \ddag}$, S. Korenblit$^1$, J. Smith$^1$, A. Lee$^1$,\\
E. E. Edwards$^1$, C.-C. J. Wang$^2$, J. K. Freericks$^2$  and C. Monroe$^1$}

\affiliation{$^1$ Joint Quantum Institute, University of Maryland Department of Physics and  
                  National Institute of Standards and Technology, College Park, MD  20742 \\
             $^2$ Department of Physics, Georgetown University , Washington, DC  20057}
\date{\today}

\begin{abstract}
Frustration, or the competition between interacting components of a network, is often responsible for 
the complexity of many body systems, from social \cite{socialnetwork} and neural \cite{nets} networks to protein folding \cite{protein} and magnetism \cite{Diep, frust, Sachdev}.  In quantum magnetic systems, frustration arises naturally from competing spin-spin interactions given by the geometry of the spin lattice or by the presence of long-range antiferromagnetic couplings.  Frustrated magnetism is a hallmark of poorly understood systems such as quantum spin liquids, spin glasses \cite{spinglass,Sachdev08} and spin ices \cite{spinice}, whose ground states are massively degenerate and can carry high degrees of quantum entanglement \cite{Nielsen04,KimNature}. The controlled study of frustrated magnetism in materials is hampered by short dynamical time scales and the presence of impurities, while numerical modeling is generally intractable when dealing with dynamics beyond $N\sim 30$ particles \cite{Sandvik}. Alternatively, a quantum simulator \cite{Feynman} can be exploited to directly engineer prescribed frustrated interactions between controlled quantum systems, and several small-scale experiments have moved in this direction \cite{NatPhysQSIM, Schaetz08, KimNature, IslamNatComm, Greiner-Simon, MaNatPhys, Bollinger}. In this article, we perform a quantum simulation of a long-range antiferromagnetic quantum Ising model with a transverse field, on a crystal of up to $N=16$ trapped \Yb atoms.  We directly control the amount of frustration by continuously tuning the range of interaction and directly measure spin correlation functions and their dynamics through spatially-resolved spin detection.  We find a pronounced dependence of the magnetic order on the amount of frustration, and extract signatures of quantum coherence in the resulting phases.
\end{abstract}

\maketitle  

Cold atoms are ideal platforms for the simulation of frustrated spin models, with the ability to tailor interactions with external fields and perform projective measurements of the individual spins \cite{NatPhysQSIM}.
Neutral atomic systems are typically limited to nearest neighbor interactions \cite{Greiner-Simon}, although geometrically-frustrated interactions can be realized in certain optical lattice geometries \cite{kagome}.  
The natural long-range interaction between cold atomic ions \cite{Porras04}
has led to the engineering of Ising couplings between individual trapped ion spins \cite{Schaetz08, KimPRL, Bollinger, Wunderlich}
and the observation of spin frustration and quantum entanglement in the smallest system of three spins \cite{KimNature}. 
In this article we implement variable-range antiferromagnetic (AFM) Ising interactions and transverse magnetic fields with up to $N=16$ atomic ion spins, using optical dipole forces.  We directly measure the emergence and frustration of magnetic order through spatially-resolved imaging of the spins, in a system that approaches a complexity level where it becomes difficult or impossible to calculate the ground state order or the spin dynamics.  

We simulate the Ising model with long range antiferromagnetic interactions, given by the Hamiltonian 
($h=1$)
\begin{eqnarray}
\label{Ham}
 H=\sum_{j<i} J_{ij}\sigma_{x}^{(i)}\sigma_{x}^{(j)}-B\sum_{i}\sigma_{y}^{(i)},
\end{eqnarray}
where $\sigma_{\gamma}^{(i)}$ $(\gamma = x,y,z)$ are the spin-$1/2$ Pauli operators for the $i$th spin 
$(i=1,2,\ldots N)$, $B$ is the effective transverse magnetic field, and $J_{ij}>0$ is the Ising coupling matrix between spins $i$ and $j$, falling off with the lattice spacing $|i-j|$ approximately as
\begin{eqnarray}
\label{Jij}
 J_{ij} \approx \frac{J_0}{|i-j|^{\alpha}},
\end{eqnarray}
where $0<\alpha<3$ \cite{Porras04}.

For $B\gg J_{ij}$ on all pairs, the spins are polarized along the effective transverse magnetic field in the ground state $\left|\uparrow_y \uparrow_y \uparrow_y \cdots \right\rangle$ 
of the Hamiltonian in Eq. \ref{Ham}, where $\upy$ denotes a spin along the $+y$-direction of the Bloch sphere. As the ratio of $B$ to the Ising couplings is reduced, the system crosses over to an ordered state dictated by the form of the Ising couplings, and the spectrum of energy levels depends on the range of the interactions.  For any finite range interaction, the staggered AFM states 
$\left|\uparrow\downarrow\uparrow\downarrow \cdots\right\rangle$ and 
$\left|\downarrow\uparrow\downarrow\uparrow \cdots\right\rangle$ constitute the doubly degenerate ground state manifold at $B=0$, with the degeneracy arising from the time reversal or the global spin flip symmetry of the Hamiltonian. Here $\up$ and $\dn$ are spins oriented along the Ising or $x$-direction of the Bloch sphere. Thus the system exhibits nearest-neighbor AFM or N\'{e}el ordering at  sufficiently low temperatures. 
When the interactions are uniform over all pairs of spins ($\alpha\to 0$), the excitation gaps close significantly with maximum frustration, leading to a finite entropy density. In this case, any spin configuration with a net magnetization of zero ($1/2$) for even (odd) numbers of spins belongs to the ground state.
\begin{figure}
\begin{center}
\includegraphics[width=1.0\linewidth]{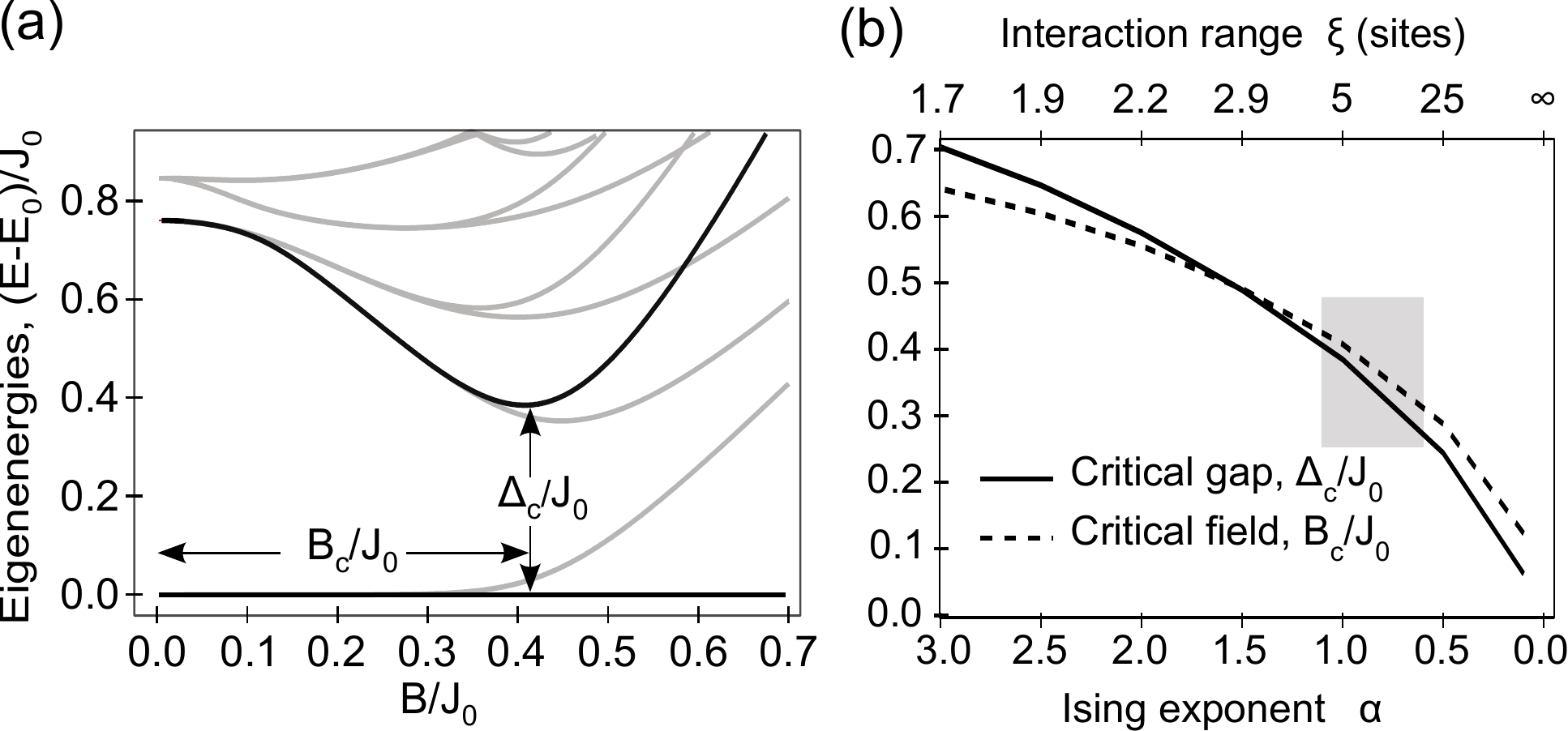}
\end{center}
\caption{Theoretical energy spectrum and critical gap in the long range antiferromagnetic Ising model (Eq. \ref{Ham}) for $N=10$ spins. {\bf{(a)}} For Ising exponent $\alpha=1$ (characteristic range of $\xi=5$ sites), there are few low-lying energy states of the frustrated Hamiltonian as a function of the dimensionless parameter $B/J_{0}$. The spacing between the ground state and the first coupled excited state reaches a bottleneck at a critical value $B_{c}/J_{0}$ with critical gap $\Delta_c$.  {\bf{(b)}} Theoretical dependence of $B_{c}/J_{0}$ (dotted line) and $\Delta_c/J_0$ (solid line) on the range of the interaction. As the interaction range increases, the competing long range couplings make it easier to create excitations and the critical gap is reduced, so a relatively small effective transverse field can break the spin ordering. Both parameters approach zero as $\alpha\to 0$ or $\xi\to\infty$. Current experiments are performed with parameters in the shaded region.} 
\label{fig:gaps}
\end{figure}

Between the limits $B=0$ and $B\gg J_{ij}$, the energy spectrum features a minimum gap, whose position and size depends on the amount of frustration in the system, or the interaction range.
(The interaction range is defined as the number of lattice spacings $\xi$ where the interaction falls off to $20\%$ of the nearest neighbor Ising coupling: $\xi=5^{1/\alpha}$.)
Figure \ref{fig:gaps}a shows a few low lying energy states of the Hamiltonian in Eq. \ref{Ham} for an interaction range of $\xi=5$ (corresponding to $\alpha=1$). The first excited eigenstate merges with the ground state for small $B/J_{0}$ and has the same spin order as the ground state near $B/J_{0}=0$. The critical gap $\Delta_c$ between the ground and the first coupled excited state determines the adiabaticity criterion \cite{EdwardsPRB}. Figure \ref{fig:gaps}b compares the position (red dotted line) and size (black solid line) of the critical gap of the Hamiltonian for various ranges.  As the range and hence the amount of frustration increases, the critical field is pushed towards zero, and the gap closes. 
The signature of frustration is the density of states near the ground state, and not merely the order in the ground state. In order to observe the effects of frustration, we therefore quench the system by ramping the effective transverse magnetic field faster than the critical gap ($|\dot{B}/B| > \Delta_c)$ to populate the lowest coupled excited states. The observed spin order depends on the resulting degree of excitation and hence on the level of frustration. 

The spins are stored in a collection of \Yb ions confined in a linear radiofrequency (Paul) trap, with the effective spin-$1/2$ system represented by two hyperfine `clock' states within each ion $\upz$ and $\dnz$, separated by the hyperfine frequency $\nu_{HF}=12.642819$ GHz \cite{OlmschenkYb}. The variable-range AFM Ising interactions are generated by applying off-resonant spin-dependent optical dipole forces \cite{KimPRL} that drive stimulated Raman transitions between the spin states while modulating the Coulomb interaction between the ions in a controlled way (see Appendix I).  The effective magnetic field is generated by simultaneously driving coherent transitions between the spin states with a $\pi/2$-phase shift with respect to the dipole force beams.  
At any time, we measure the state of the spins by illuminating the ions with resonant radiation and collecting state-dependent fluorescence on an imager with site-resolving optics \cite{OlmschenkYb}.  From this information we can extract all spin correlation functions (see Appendix II).

\begin{figure}
\begin{center}
\includegraphics[width=0.9\linewidth]{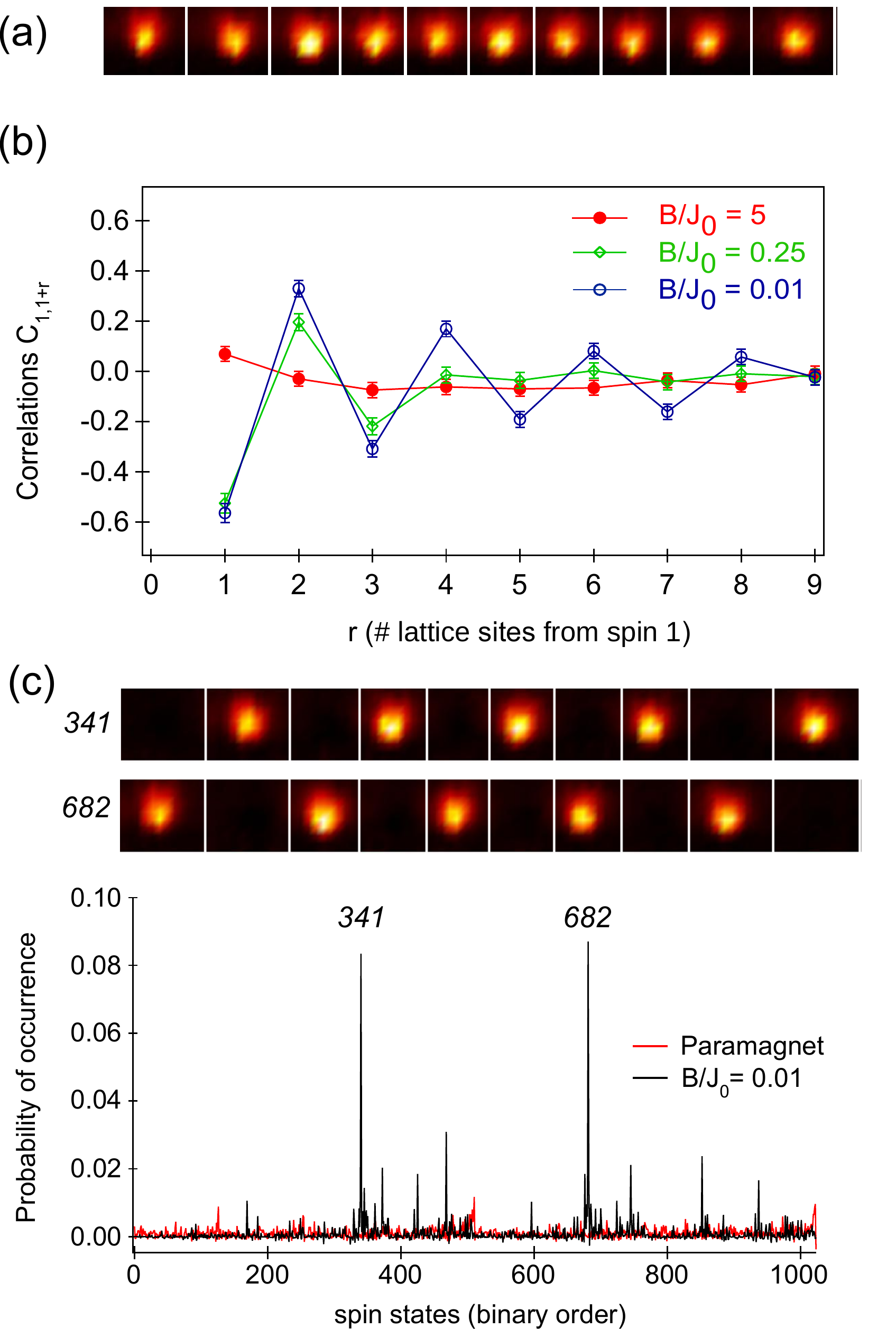}
\end{center}
\caption{Quantum phase transition from a paramagnet to an antiferromagnet (Eq. \ref{Ham}), with $J_{ij}\sim |i-j|^{-1.1}$ in a system of $10$ spins.  {\bf{(a)}} Image of $10$ trapped ions, with a distance of $22 \mu$m between the first and last ion. {\bf{(b)}} Measured two-point correlation function between a chosen spin (on the edge) and the other spins separated by $r$ lattice sites, $C_{1,1+r}=\langle\sigma_{x}^{(1)} \sigma_{x}^{(1+r)}\rangle-\langle\sigma_{x}^{(1)}\rangle\langle \sigma_{x}^{(1+r)}\rangle$, averaged over 4000 experiments for each value of the dimensionless parameter $B/J_{0}$. For $B/J_{0}=5$ the spins are initially polarized along the transverse $y$-field with little correlation along the Ising $x$-direction. As the field is reduced the spins crossover to predominantly AFM states 
$\left|\uparrow\downarrow\uparrow\downarrow \cdots\right\rangle$ and 
$\left|\downarrow\uparrow\downarrow\uparrow \cdots\right\rangle$, resulting in alternating signs in the two point correlations with separation. {\bf{(c)}} Measured occurance probability of all $2^{10}=1024$ states at $B/J_0\approx 5$ (paramagnetic state, red trace) and $B/J_0\approx 0.01$ (AFM phase, black trace). The states are listed in binary order, with spin $\dn \equiv 0$ and $\up \equiv 1$.
The residual peaks in the red trace are consistent with detection errors biased toward states with many $\up$ spins such as $127$, $255$, 
and $1023$. The two tall peaks in the black trace at $341$ and $682$ correspond to two N\'{e}el-ordered staggered AFM states, shown with camera images of these cases and contributing $\sim17\%$ to the population.} 
\label{fig:AFMcorrvstime}
\end{figure}

The quantum simulation begins by optically pumping all spins to the $\dnz$ state and then coherently rotating each spin about the $x-$axis of the Bloch sphere to initialize each spin in state $\upy$ along the effective transverse magnetic field. The Hamiltonian (Eq. \ref{Ham}) is then switched on with an initial field $B_{0}\approx 5 J_{0}$, where $J_{0}$ is the average nearest neighbor Ising coupling, thus preparing the spins in the ground state of the initial Hamiltonian with a fidelity better than $97\%$.  The effective magnetic field is ramped down exponentially in time with time constant $400$ $\mu$s (and no longer than $2.4$ ms overall) to a final value $B$ of the transverse field. At this point the Hamiltonian is switched off, freezing the spins for measurement.  We finally measure the $x-$ or $y-$component of each spin $\langle \sigma_x^{(i)} \rangle$ or $\langle \sigma_y^{(i)} \rangle$ by first rotating our measurement axes with an appropriate global $\pi/2$ pulse similar to the initialization procedure, before capturing the spin-dependent fluorescence on the imager. The experiments are repeated $\sim 2000$-$4000$ times to collect statistics on the resulting state. 

From these measurements, we can construct various order parameters appropriate for observing low energy AFM states. 
The staggered magnetization $m_{s}=\frac{1}{N}|\sum_{i=1}^{N}(-1)^i \langle \sigma_{x}^{(i)}\rangle|$ 
delineates paramagnetic and AFM order, and also quantifies spin flip excitations. 
The fourth moment of this magnetization, known as a Binder cumulant 
$\overline{g}_{s}=3/2 - \overline{m_{s}^4}/2(\overline{m_s^2})^2$ varies from $0$ to $1$ 
as the paramagnetic state gives way to AFM order and is scaled to remove finite size effects, 
and is averaged over independent experimental realizations.
We also can form any correlation function of the spins such as the two-point correlation  
$C_{i,j} = \langle \sigma_{x}^{(i)} \sigma_{x}^{(j)} \rangle - \langle \sigma_{x}^{(i)} \rangle \langle \sigma_{x}^{(j)}\rangle$, allowing a direct probe of spin order for each experimental realization. The Fourier transform of this correlation function 
is the structure function $S(k)=\frac{1}{N-1} |\sum_{r=1}^{N-1} C(r)e^{ikr}|$, 
where $C(r) = \frac{1}{N-r}\sum_{m=1}^{N-r}C_{m,m+r}$ is the average correlation over $r$ sites in the chain.  The structure function shows spin order versus wavenumber $k$, with $S(k=\pi)$ singling out 
the presence of the nearest neighbor N\'{e}el AFM order. 

Figure \ref{fig:AFMcorrvstime} shows the onset of antiferromagnetic ordering in the quantum simulation of the frustrated transverse field Ising model in a system of 10 spins. Two-point spin correlations $C_{1,r}$ between a chosen edge spin and the other spins are presented in Fig. \ref{fig:AFMcorrvstime}b at various stages in the ramp $B/J_0=5, 0.25$, and $0.01$. For larger transverse magnetic fields, there are no appreciable correlations between the spin components along the Ising direction. As the ratio of $B/J_0$ is lowered however, a zig-zag pattern emerges, with negative (positive) correlations between spins separated by odd (even) lattice spacings. For $B/J_0\approx 0.01$ the nearest-neighbor spin correlation reaches about $60\%$, and the correlation length (defined to be the distance at which the absolute correlation drops to $1/e$ of the nearest neighbor value) reaches about $6$ lattice sites. The effective field was ramped exponentially down from $B/J_0\approx 5$ with a time constant of $400$ $\mu$s in this experiment. This ramping is not slow enough to be adiabatic, and the diabatic effects prevent the spin ordering to reach a perfect AFM phase. Figure \ref{fig:AFMcorrvstime}c shows the measured probabilities of all 
$2^{10} =1024$ possible spin states measured along the Ising $x$-direction, sorted in binary order with spin $\dn \equiv 0$ and $\up \equiv 1$.  
Since the detection fidelity of each spin within a chain is $\sim93\%$, the probability of correctly detecting a particular state of the 
$10$ spins is only $0.93^{10}\approx 48\%$.  We recover effective detection efficiencies of $\sim 98\%$ per spin by post-filtering the measurements based on calibrating the known detection errors for each spin \cite{DuanDetection}, as described in Appendix II.
The initial paramagnetic phase shown in red ($B/J_0\approx 5$) exhibits a roughly uniform probability of $1/1024\approx 0.1\%$ for each state (the residual peaks in the red trace are consistent with detection errors).
The spin ordered phase shown in black ($B/J_0 = 0.01$) displays the emergence of the two AFM states, each with an occupation probability of about $8.5\%$. Other prominent peaks correspond to single spin-flips and other low-lying excitations from the two ground states. 

\begin{figure}
\begin{center}
\includegraphics[width=1.0\linewidth]{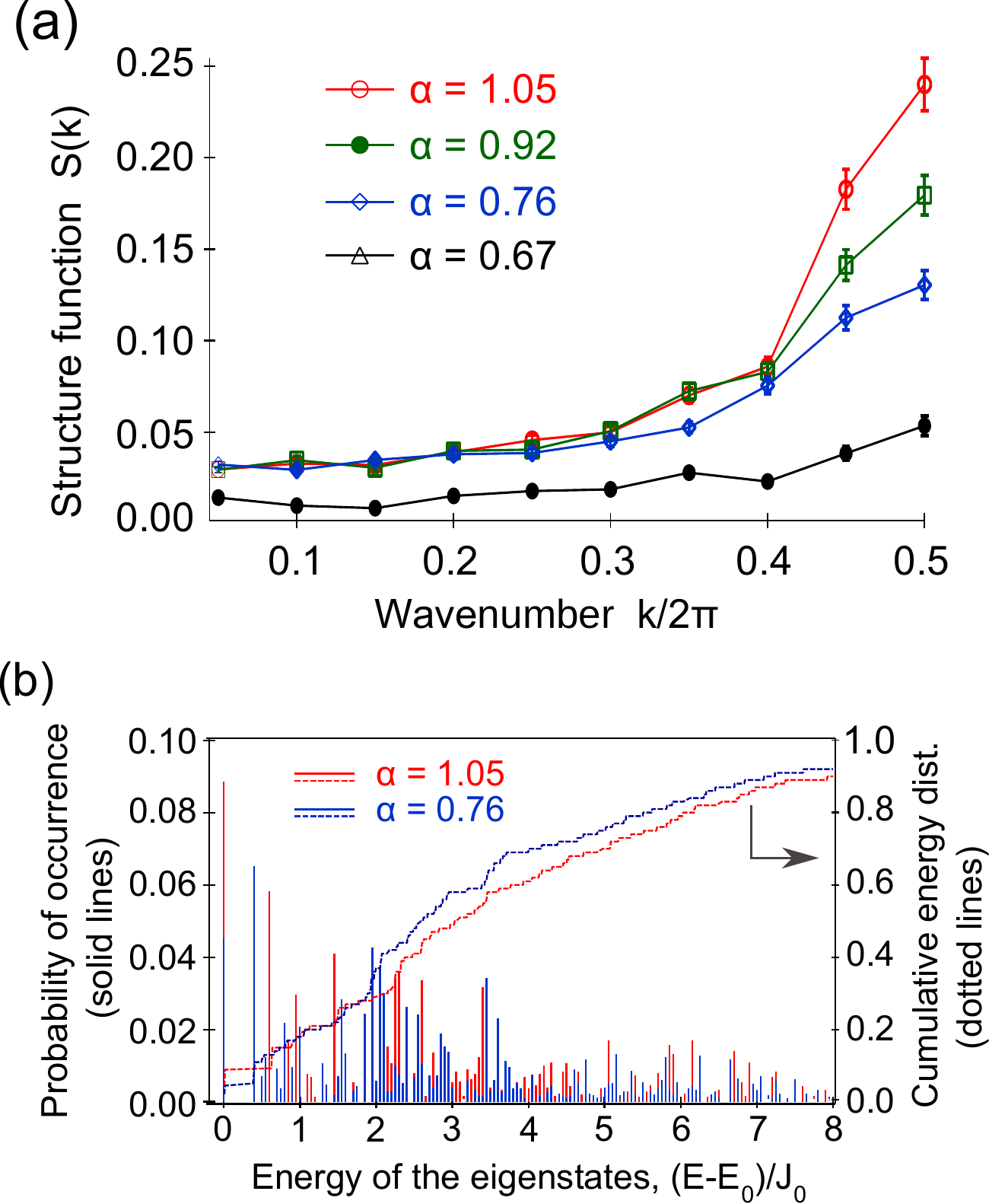}
\end{center}
\caption{Spin order vs the range of the Ising couplings for $B/J_0\approx 0.01$. {\bf{(a)}} Structure function $S(k)$ vs wavevector $k$ for various ranges of AFM interactions. As the range of interaction increases, the growing frustration suppresses the ground state order in the quantum simulation, captured by the decreasing value of $S(k=\pi)$. The error bars include statistical fluctuations and estimated detection uncertainties. The detection errors are larger for the longest range of interactions, owing to spatial crosstalk from their closer spacing.  {\bf{(b)}} Distribution of observed states in the spin system, sorted according to their energy $E_i$ that was previously calculated by diagonalizing Eq. \ref{Ham} with $B=0$.  Data is presented for two ranges (red for $\alpha=1.05$ and blue for $\alpha=0.76$).  The dashed lines indicate the cumulative energy distribution functions for these two ranges.} 
\label{fig:10_ion_structure_function}
\end{figure}
In Fig. \ref{fig:10_ion_structure_function} we probe the frustration in the system for various ranges of interactions. Here we look at the spin order achieved in the quantum simulation when the external magnetic field is ramped down to $B/J_0 \approx 0.01$ for four different ranges of interactions. In Fig. \ref{fig:10_ion_structure_function}a we show the measured structure function $S(k)$ at wavevector values $k=\frac{\pi}{10},\frac{2\pi}{10},...,\pi$ from the measured two-point correlation functions. To directly compare the different interaction ranges, we choose the same external magnetic field ramp time constant, $\tau=0.4/J_0$. As the range of interaction increases, the ground state AFM order (given by the structure function peak at $k=\pi$) disappears, denoting increased occupation of the excited states as the frustration grows. Figure \ref{fig:10_ion_structure_function}b displays the observed distribution of energy $P(E_i)$ for $\alpha=1.05$ (shorter range) and $\alpha=0.76$ (longer range) power law exponents, along with the cumulative energy distribution function.
The eigenenergies of each configuration are calculated using Eq. \ref{Ham} with $B=0$. Note that for the longer range interactions, excitations are more prevalent, and the energy gap between the ground and the first excited state is reduced, both signatures of increasing frustration in the system.  
The observed final entropy per particle $S = -\frac{1}{N}\sum_i P(E_i) \text{log} P(E_i)$
is seen to increase as the interaction range grows, from $S(\alpha=1.05) = 0.832$ to 
$S(\alpha=0.67) = 0.903$, which is also a signature of the increased frustration in the system.  As a reference, the paramagnetic state distribution shows an entropy per particle of $0.959$, which is slightly less than unity due to detection errors.
\begin{figure}
\begin{center}
\includegraphics[width=1.0\linewidth]{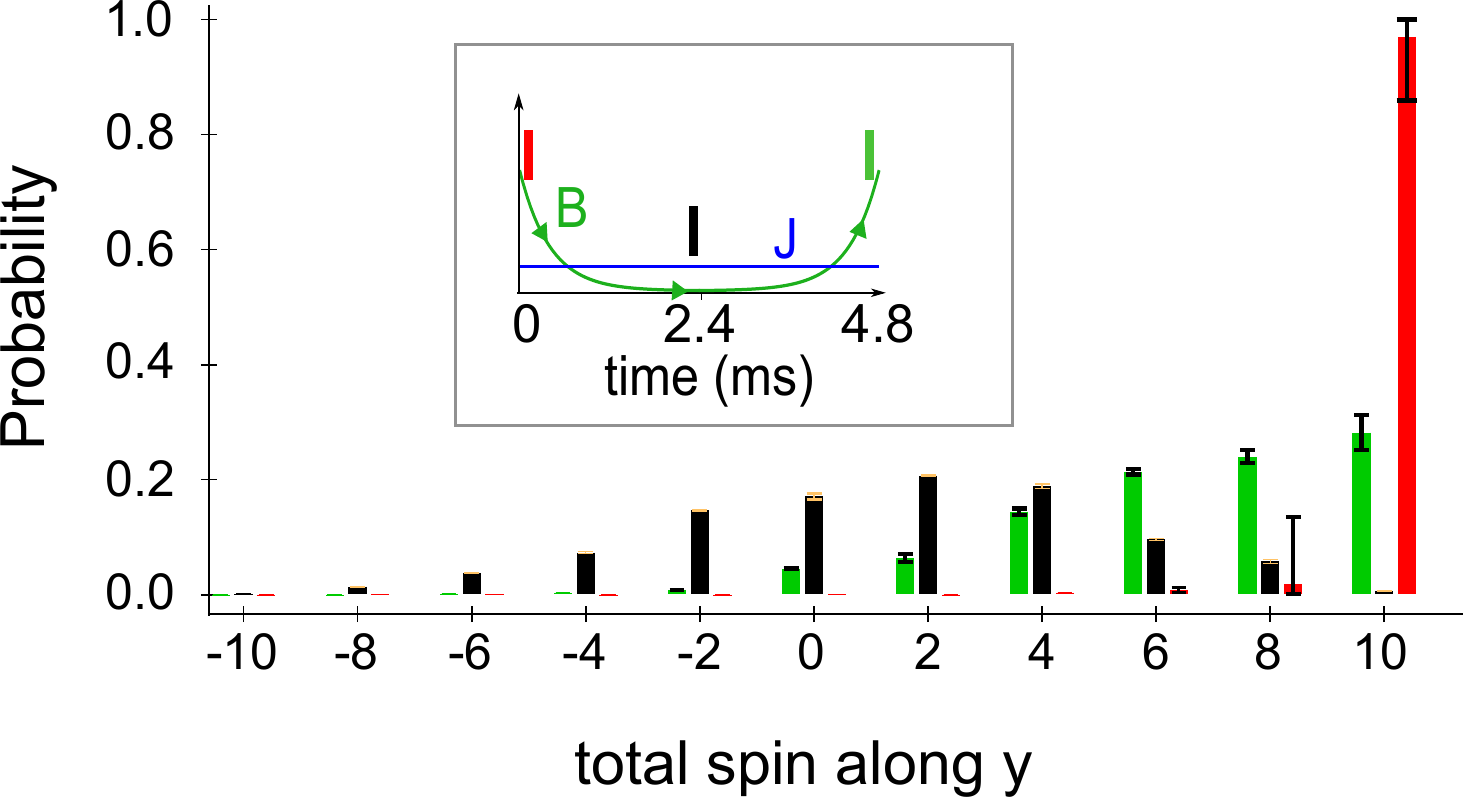}
\end{center}
\caption{Quantum coherence in the simulation. Probabilities of different values of the total spin component along $y$ in the initial polarized state (red), when the transverse field is ramped to $\approx 0.01 J_0$ (black), and when the transverse field is reversed back to its initial value (green). After reversal, the y-magnetization 
returns to $\sim 63\%$ of its initial value, indicating quantum coherence in the evolution. The trajectory of the transverse field ($B$, in green) and all the Ising couplings ($J$, in blue) is shown in the inset. } 
\label{fig:10_ion_coherence_histogram}
\end{figure}

The above measurements of the state distribution concern only the diagonal components or populations of the density matrix. To characterize the quantum coherence in the simulation, we retrace the external magnetic field back to its initial value after ramping it down to almost zero, and measure each spin along the transverse $(y)$ magnetic field. Figure \ref{fig:10_ion_coherence_histogram} shows the distribution of the total transverse spin
$S_y = \sum_i \langle \sigma_y^{(i)} \rangle$ at three different times: first at the beginning of the simulation, second when the transverse field has been ramped down to nearly zero, and third after the transverse field returns to its initial value. The initial state is ideally a delta function at $S_y=10$, but finite detection efficiency broadens the distribution. At the lowest value of the field ($B/J_0\approx 0.01$), the transverse magnetization is distributed near $S_y=0$, as the spins are presumably ordered along the Ising $x$-direction. When the external field is ramped back to its initial state, the distribution of the total spin returns toward the initial distribution, with a magnetization that is approximately $63\%$ of its initial value, indicating that quantum coherence is maintained throughout the simulation.

\begin{figure}
\begin{center}
\includegraphics[width=1.0\linewidth]{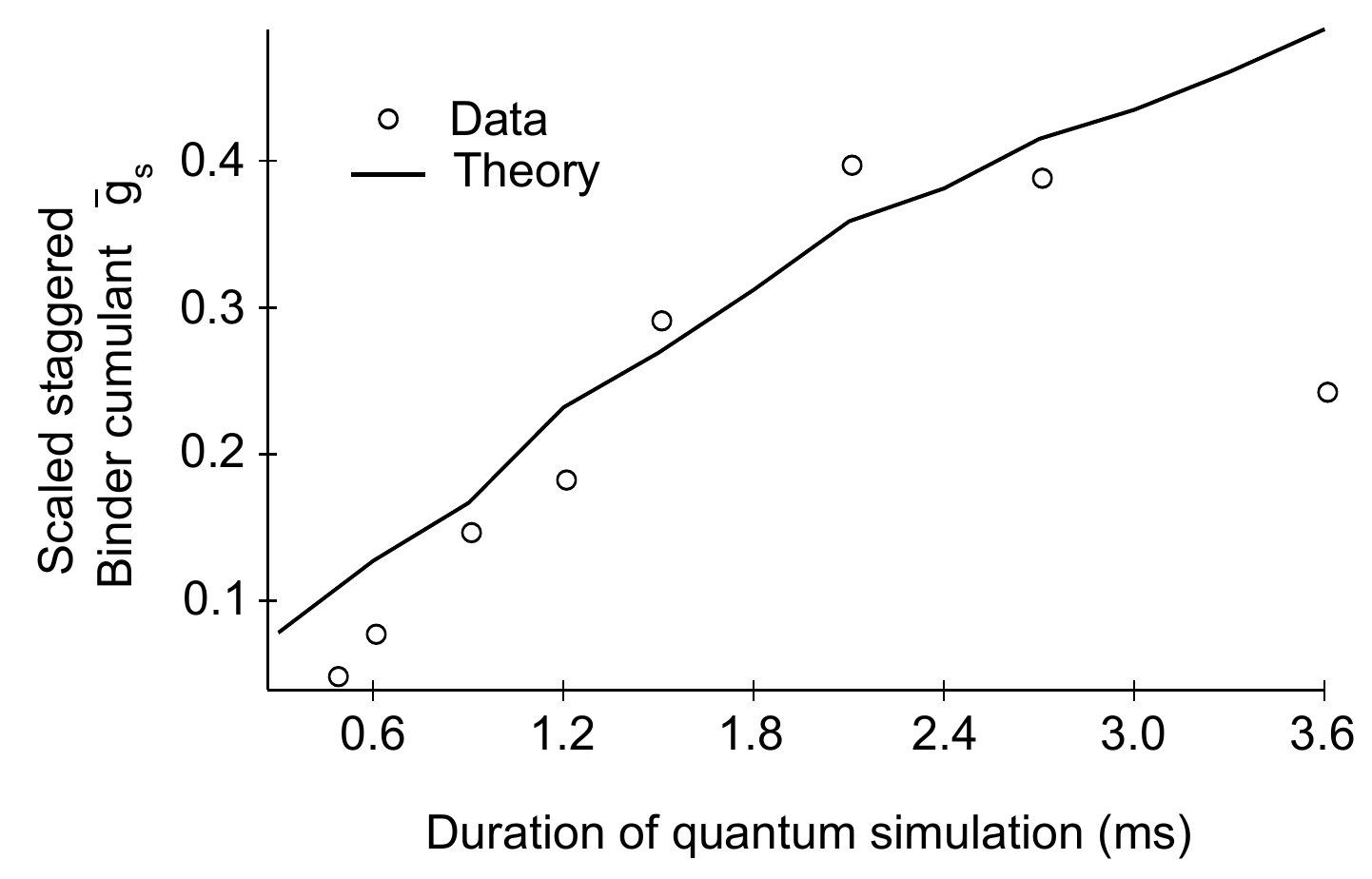}
\end{center}
\caption{Spin order vs speed of ramp. The spins are initialized with $B/J_0=5$ and the transverse field is ramped exponentially down for $6$ time constants, and the experiment is repeated for various values of time constant $\tau$. Here we plot the scaled staggered Binder cumulant, $\bar{g_{s}}$ versus the total duration $6\tau$ of the ramp, and the solid line indicates the theoretical expectation of how the Binder 
order parameter should grow with time.  As the Hamiltonian evolves more slowly, the observed spin order shows more ground state order, and less excitation for ramp times under $\sim 2.5 $ms. However, for longer times, the spins become disordered, implying external decoherence in the system.}
\label{fig:orderVsTau}
\end{figure}
In order to probe decoherence in the simulation, we repeat the experiment with various ramping speeds of the effective magnetic field. In Fig. \ref{fig:orderVsTau} we plot the AFM order parameter $\bar{g_{s}}$ vs the total duration for the experiment for a long range coupling ($\alpha=1.12$ power law with distance) for $N=10$ spins. Each data point represents the spin order achieved after ramping the magnetic field $B$ down exponentially from $5 J_0$ for a total duration of $6$ time constants. The AFM order grows with slower ramping, as expected, for up to $\tau=400\mu s$. But we also observe a saturation and then decay in the spin order, which might indicate the presence of decoherence in the system at long times.  During the simulation, spontaneous Raman scattering from the optical beams is expected to occur at a rate of less than $10^{-5}\Omega \sim 6$ $s^{-1}$ per spin \cite{CampbellPRL10}, which is consistent with separate measurements of the spin relaxation from a single spin and is therefore not expected to contribute to decoherence given the timescales in the experiment. The phonon population is expected to be well under $10\%$ for all the data presented here \cite{phonons}. A principal source of decoherence appears to be the intensity fluctuations in the Raman beams, due to beam pointing instabilities, and fluctuations in the optical power. 

\begin{figure}
\begin{center}
\includegraphics[width=1.0\linewidth]{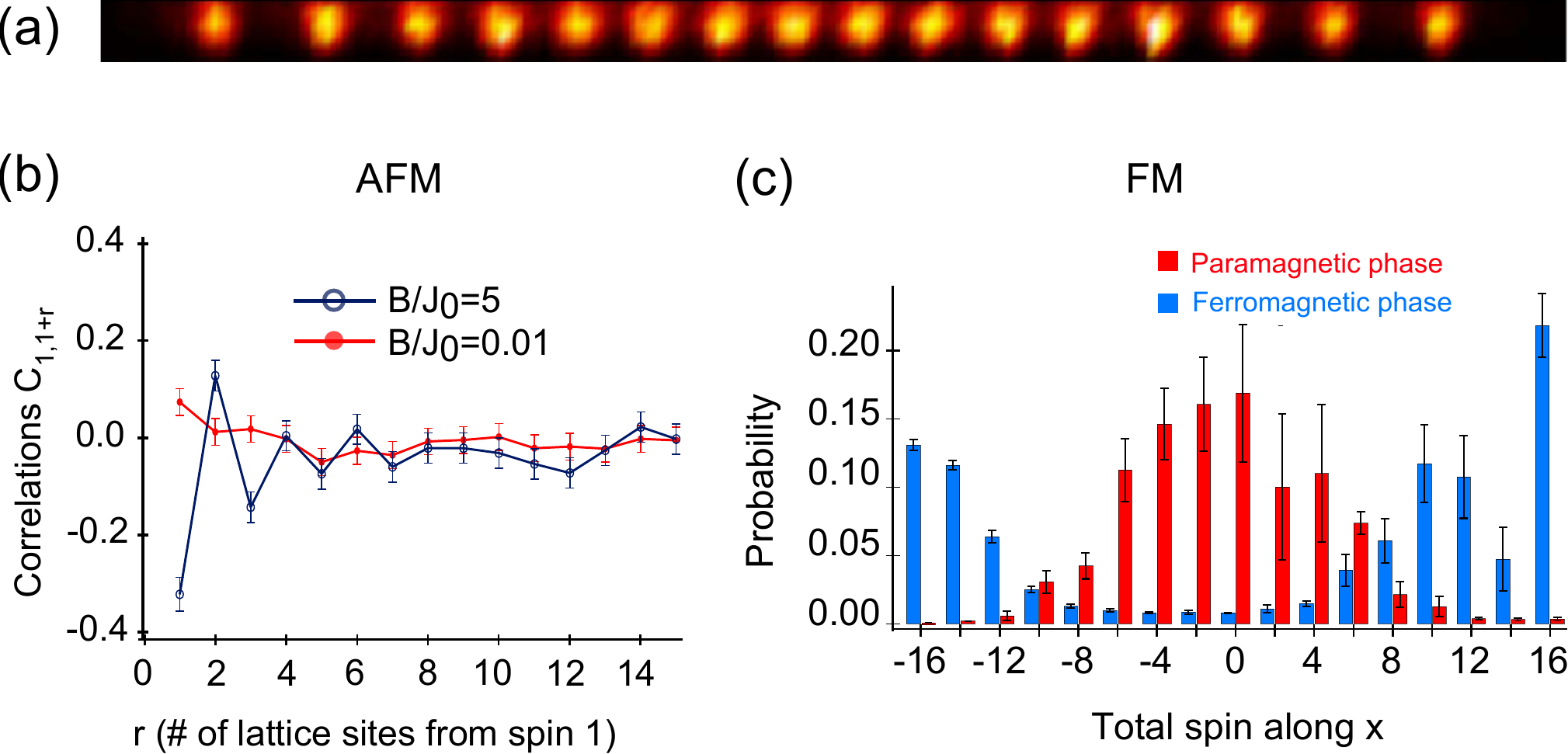}
\end{center}
\caption{Magnetic ordering in $N=16$ spins. \textbf{a.} Image of $16$ trapped ions, with a distance of $30\mu$m between the first and last ion.  \textbf{b.} Pair correlation function measured at various stages of the quantum simulation, for $B/J_0 = 5$ (red) and $B/J_0 = 0.01$ (blue) in an AFM coupling falling off with distance as $J_{ij}\sim |i-j|^{-1}$ among $N=16$ spins. While small amounts of staggered order are seen, it is tempered by the small gaps and frustration in the low energy states. \textbf{c.} In contrast, for all FM couplings (again with $J_{ij}\sim |i-j|^{-1}$) the gaps are large and clear FM order is seen. Here the measured distribition of magnetization is plotted.  The paramagnetic phase of $16$ spins is indicated in red, and after the field is ramped to nearly zero, the distribution clearly bifurcates, indicating population weighted heavily towards the FM states $\left|\downarrow\downarrow\downarrow\cdots\right\rangle$ and 
$\left|\uparrow\uparrow\uparrow\cdots\right\rangle$. The resulting magnitude magnetization is approximately $73\%$.} 
\label{fig:16_ion}
\end{figure}
Simulating adiabatic evolution in frustrated models is generally slower than simulations without frustration, because the relevant energy gaps are smaller.  As a direct comparison of these cases, we compare the quantum simulation of the long range AFM and ferromagnetic (FM) Ising models in a system of $N=16$ spins, shown in Fig. \ref{fig:16_ion}.  For the FM experiment (Fig. \ref{fig:16_ion}c), we initialize the spins in the \textit{highest} energy state with respect to the transverse field $|\!\downarrow_{y}\downarrow_{y}\downarrow_{y} \cdots \rangle$ and ramp the field down as before.  For the same simulation speed, we find that the best AFM nearest-neighbor correlation (Fig. \ref{fig:16_ion}b) is only $\sim30\%$, corresponding to a staggered magnetization of about $30\%$, while the simulation of the ferromagnetic model shows a clear ferromagnetic spin order across the chain, reaching $\sim 73\%$ magnetization.  We have also observed a level of $74\%$ FM magnetization emerging in $N=18$ spins. The current limit on calculating dynamics of fully connected spin models is approximately $N=30$ spins, a size that may be reached with technical upgrades in the hardware, including lower vacuum chamber pressures to prevent collisions with the background gas, better stability of the optical intensities, and higher optical power so that fluctuations in the beam inhomogeneities can be suppressed. 

\section*{Appendix I: Generating variable-range AFM Ising interactions} 

The Ising interaction is generated by globally addressing the ions with two 
off-resonant laser beams at $\lambda = 355$ nm \cite{CampbellPRL10}, intersecting at right angles with wavevector difference $\Delta k$ along a principal axis of transverse ion motion \cite{ZhuTransverse}.
These beams have beatnote frequencies $\nu_{HF}\pm\mu$ which drive stimulated Raman transitions near the upper and lower motional sidebands of transverse motion in order to impart a spin-dependent optical dipole force \cite{DidiRMP}. By setting the beatnotes sufficiently far from the sidebands, motional excitations can be made negligible, resulting in a nearly pure spin-spin coupling mediated by the Coulomb interaction \cite{KimPRL}.
The effective transverse magnetic field is generated by simultaneously driving a resonant stimulated Raman transition between the spin states with a beatnote frequency $\nu_{HF}$ and a phase that is shifted by $\pi/2$ with respect to the mean phase of the sideband fields.  
The resulting Ising coupling matrix $J_{ij}$ is given by a sum over contributions from each normal modes of collective motion at frequency $\nu_m$, 
\begin{equation}
J_{ij} = \Omega^2 \nu_{R} \sum_{m=1}^{N}\frac{b_{i,m}b_{j,m}}{\mu^2 - \nu_m^2}
\label{Jij_full}
\end{equation}
where $\nu_R = h/M\lambda^2 = 18.5$ kHz is the recoil frequency associated with the dipole force, $M$ is the mass of a single ion, $b_{i,m}$ is the orthonormal mode component of ion $i$ with mode $m$, and $\Omega$ is the (uniform) single spin flip Rabi frequency, proportional to the laser intensity at each ion. The symmetric detuning $\mu$ of the beatnote from the spin-flip transition controls the sign and range of the interactions \cite{KimPRL}. When $\mu$ is set larger than the highest (center-of-mass or COM) mode frequency $\nu_1$, every interaction is AFM, and we can empirically approximate Eq. \ref{Jij_full} as falling off with distance as a power law $J_{ij}\approx J_{0}/|i-j|^{\alpha}$ with $0 < \alpha < 3$ \cite{Porras04} and $J_0 \propto 1/N$. While the COM mode mediates a uniform interaction between all pairs of spins, the other modes introduce non-uniformity in the interactions, and effectively reduce the range of AFM interaction.  In practice, we control the interaction range by changing the bandwidth of the transverse mode spectrum, achieved by varying the axial confinement of the ions in the Paul trap. The Ising couplings $J_{ij}$ depend not only on the spatial separation $|i-j|$, but also on the site $i$ itself due to the finite size of the system, with $\sim 10\%$ inhomogeneities across the chain. We average over all the couplings between spins separated by a given number of lattice sites to estimate the power law range exponent $\alpha$ in Eq. \ref{Jij} \cite{Bollinger}. 

In the experiment we use global Raman beams each with an optical power of $\sim 1$W, having horizontal and vertical waists of $\sim 150$ $\mu$m and $\sim 7$ $\mu$m respectively to address the ions. This produces a spin-flip Rabi frequency $\Omega\sim 600$ kHz on resonance, with less than $5\%$ inhomogeneity across the chain. We set the beatnote detuning to $\mu \approx \nu_{1}+3\eta\Omega$, where $\eta = \sqrt{\nu_R/\nu_1}$ is the single ion Lamb-Dicke parameter.  This keeps the (primarily COM) phonon excitation probabilities sufficiently low for any setting of the range. The typical nearest neighbor Ising coupling is $J_0 \sim 1$ kHz for $N=10$ spins.
In principle, the Ising interaction range can be varied from uniform to dipolar ($0<\alpha<3$), but in this experiment the 
axial frequencies was only varied between 0.62 MHz and 0.95 MHz, and given the COM transverse frequency of $\nu_1 = 4.1$ MHz, this results in a range of Ising power-law exponents $0.7<\alpha<1.2$, or a variation of the range of interactions between $\xi=4$ to $\xi=10$ sites. 

\section*{Appendix II: Accounting for finite detection efficiency of N-particle correlations}
We detect the spin states using spin dependent fluorescence collected through f/2.1 optics on an intensified charge-coupled-device (ICCD) camera or a photomultiplier tube (PMT). The spin state $\upz$ fluoresces from the near resonant detection beam, and appears bright, while the spin state $\dnz$ scatters little from the off-resonant detection beam, thus appearing dark. The imager has single site resolution, allowing us to directly measure the two point correlations to probe the AFM order. The single spin detection efficiency is about $\epsilon=93\%$ on the ICCD imager, and about $98\%$ on the PMT, the reduced efficiency on the ICCD being due to electronic and readout noise. 
To account for the spatial overlap of the fluorescence from neighboring bright ions, we fit each single shot image in the experiment to a sum of $N$ Gaussians, where $N$ is the number of ions. The center and width of the Gaussians are pre-calibrated from images of all spins prepared in the bright states, with background subtraction from all spins prepared in the dark states. 
The probability of correctly identifying a $N=10$-body spin state is only $\epsilon^N\approx 48\%$, so we post process the detected states to account for this finite detection efficiency \cite{DuanDetection}. The probability of incorrectly assigning an $N$-qubit state $\ket{\,i}$ to the actual underlying state $\ket{\,j}$ is $M_{ij}=(1-\epsilon)^{\beta_{ij}}\epsilon^{N-\beta_{ij}}$, where 
$\beta_{ij}$ is the number of positions that the $N$-qubit state $\ket{\,j}$ differs from $\ket{\,i}$ through bit flips. The observed probability distribution of all $2^N$ states is given by $P'_{i}=\sum_j M_{ij}P_j$, where $P_j$ is the underlying actual distribution of states, which can be obtained by 
simply inverting the matrix $M_{ij}$ and forming $P_{i}=\sum_j M^{-1}_{ij}P'_j$. This increases the effective detection fidelity to about $98\%$, equivalent to that with the PMT. Some entries of the post-processed probabilities are slightly negative, due to fluctuations in absolute fluorescence levels that impact the values in the matrix $M_{ij}$ during measurement. 


\subsection*{Acknowledgements}  
\noindent
\dag Current address, University of California Los Angeles, Department of Physics, Los Angeles, CA  90095.

\noindent
\ddag Current address, Harvard University Department of Physics, Cambridge, MA  02138.

We thank Eugene Demler, Luming Duan, David Huse, Kihwan Kim, Philip Richerme, Rajdeep Sensarma, and Peter Zoller for critical discussions. This work is supported by the U.S. Army Research Office (ARO) Award W911NF0710576 with funds from the DARPA Optical Lattice Emulator Program, ARO award W911NF0410234 with funds from IARPA, and the NSF Physics Frontier Center at JQI.  J.K.F. was supported by the McDevitt bequest at Georgetown.

\bibliography{QSim_AFM}

\end{document}